\documentclass[10pt,conference]{IEEEtran}
\IEEEoverridecommandlockouts
\RequirePackage{fix-cm}
\usepackage[normalem]{ulem}
\usepackage{graphicx,subfigure}
\usepackage{amsmath,amssymb,amsfonts}
\usepackage{multirow}
\usepackage{lscape}
\usepackage{longtable}
\usepackage{supertabular}
\usepackage{subfigure}
\usepackage{enumitem}
\usepackage{algorithmic}
\usepackage{textcomp}
\usepackage{xcolor}
\usepackage{pgfplots}
\pgfplotsset{compat=newest}
\usepackage{caption}
\usepackage{pgfplotstable}
\usepackage{comment}
\usepackage{booktabs}
\usepackage{tcolorbox}
\usepackage{url}
\usepackage{balance}
\usepackage{color, colortbl}
\usepackage{tabularx}
\usepackage{adjustbox}
\usepackage{dirtytalk}
\usepackage{csquotes}
\usepackage{ulem}

\usepackage{pifont}

\usepackage[colorlinks = true,
            linkcolor = black,
            urlcolor  = blue,
            citecolor = black,
            anchorcolor = black]{hyperref}
\definecolor{gray}{gray}{0.8}
\definecolor{cyan}{rgb}{0.88,1,1}

\usepackage{cite}
\usepackage{amsmath,amssymb,amsfonts}
\usepackage{algorithmic}
\usepackage{graphicx}
\usepackage{textcomp}
\usepackage{xcolor}
\usepackage{tcolorbox}
\usepackage{booktabs}

\newcommand{\Maven}{Maven}
\newcommand{\npm}{npm}

\newcommand{\Bower}{Bower}

\newcommand{\Go}{Go}
\newcommand{\RubyGems}{RubyGems}

\newcommand{\PyPI}{PyPI}
\newcommand{\CRAN}{CRAN}

\newcommand{\Packagist}{Packagist}

\begin{document}
\begin{sloppy}

\title{Contrasting Third-Party \\ Package Management User Experience}


\author{\IEEEauthorblockN{Syful Islam, Raula Gaikovina Kula, Christoph Treude\textsuperscript{\textdagger}, Bodin Chinthanet,  \\ Takashi Ishio, Kenichi Matsumoto
}
\IEEEauthorblockA{\textit{Nara Institute of Science and Technology (NAIST), Japan} \\
\textit{\textsuperscript{\textdagger}University of Adelaide, Australia} \\
Email: \{islam.syful.il4, raula-k, bodin.chinthanet.ay1, ishio, matumoto\}@is.naist.jp, christoph.treude@adelaide.edu.au\textsuperscript{\textdagger}
}
}
\maketitle

\begin{abstract}
The management of third-party package dependencies is crucial to most technology stacks, with package managers acting as brokers to ensure that a verified package is correctly installed, configured, or removed from an application.
Diversity in technology stacks has led to dozens of package ecosystems with their own management features.
While recent studies have shown that developers struggle to migrate their dependencies, the common assumption is that package ecosystems are used without any issue. 
In this study, we explore 13 package ecosystems to understand whether their features correlate with the experience of their users.
By studying experience through the questions that developers ask on the question-and-answer site Stack Overflow, we find that developer questions are  grouped into three themes (i.e., Package management, Input-Output, and Package Usage).
Our preliminary analysis indicates that specific features are correlated with the user experience.
Our work lays out future directions to investigate the trade-offs involved in designing the ideal package ecosystem.
\end{abstract}


\section{Introduction}
Package management is crucial to most technology stacks in software development, especially when developers are building web or mobile applications.
Package managers serve more than five million open source packages that developers can easily adopt to introduce new functionality into their projects, without the need to create these functions from scratch\footnote{According to \url{www.libraries.io}.}.
In 2020, GitHub showed its investment in package support when it acquired the Node.js package ecosystem (i.e., {\npm}), which serves over 1.3 million packages to roughly 12 million developers, and is constantly growing each day \cite{Web:npmStat}.

Package management is an automated solution for applications that heavily rely on package dependencies. 
They solve the `dependency hell'\footnote{A term made popular by this blog \url{https://web.archive.org/web/20150708101023/http://archive09.linux.com/feature/155922}} dilemma, which is a colloquial term used to describe the frustration of users when an application has dependencies that do not run with other versions of dependencies (i.e., incompatibility).
As the number of dependencies grows (i.e., forming a large tree of interdependent packages) within the application, so do the chances of incompatibility between dependencies.
To address such problems, the package managers act as intermediary brokers between an application and a package dependency to ensure that a verified package is correctly installed, configured, or removed from an application.

Diversity in technology stacks and programming languages has led to a variety of managers that cover different package ecosystems. 
For instance, npm brokers packages that run in the Node.js environment and are written in JavaScript, while the PyPI package ecosystem is built specifically to handle Python package dependencies.
Although recent studies have investigated dependency management from the perspectives of updates (i.e., update an existing dependency to a more recent version) and migration (i.e., replace, remove, or add a new dependency)
\cite{Cox-ICSE2015,bogart2016break,kula2018developers,decan2018impact,decan2019empirical}, the focus has not been on the package manager itself.
For example, Bogart et al.~\cite{bogart2016break} cite reasons why developers do not update, including ecosystem community values such as policies, supporting infrastructure, and accepted trade-offs to negotiate dependency changes. 
Other work \cite{kula2018developers} found that 69\% of the developers claimed to be unaware of the need to update and were not likely to prioritize a library update, as it was perceived to be extra workload and responsibility.
Bogart et al.~\cite{10.1109/ASEW.2015.21} argued that awareness mechanisms based on various notions of stability can enable developers to make migration decisions.
While these studies have shown that developers struggle to migrate their dependent packages, the common assumption is that the management systems all work in the same manner. 

\begin{table*}[]
\caption{Summary of each package ecosystem and their features.}
 \label{tab:PM_features}
 \centering
\resizebox{.7\textwidth}{!}{%
\begin{tabular}{@{}llrlllr@{}}
\toprule
\begin{tabular}[c]{@{}l@{}}Package \\ Ecosystem\end{tabular} & \begin{tabular}[c]{@{}l@{}}Programming \\ Language\end{tabular} & Tiobe Rank & Environment & \begin{tabular}[c]{@{}l@{}}Dependency \\ Tree\end{tabular} & \begin{tabular}[c]{@{}l@{}}Package   \\Archive link\end{tabular} & \begin{tabular}[c]{@{}r@{}}\# of packages\\ in ecosystem\end{tabular} \\ \midrule
PyPI & Python & 2 & Python & Flat & pypi.org & 372,334 \\
Maven & Java & 3 & JVM & Flat & Maven.org & 417,669 \\
Bower & JavaScript & 7 & Node.js & Flat & bower.io & 69,625 \\
Meteor & JavaScript & 7 & Node.js & Nested & atmospherejs.com & 13,410 \\
npm & JavaScript & 7 & Node.js & Nested (v2) & npmjs.com & 1,866,208 \\
Packagist & PHP & 8 & PHP & Flat & packagist.org & 316,855 \\
Puppet & Ruby & 13 & Ruby MRI & Flat & forge.puppet.com & 6,923 \\
RubyGems & Ruby & 13 & Ruby MRI & Flat & rubygems.org & 173,603 \\
CRAN & R & 14 & RStudio & Flat & cran.r-project.org & 20,324 \\
CPAN & Perl & 15 & Perl & Flat & metacpan.org & 38,459 \\
GO & Golang & 20 & Go & Flat & pkg.go.dev & 390,438 \\
NuGet & C\#, VB & 5, 6 & .NET & Flat & nuget.org & 264,221 \\
Anaconda & Python, R, C\# & 2, 14, 5 & Anaconda & Flat & anaconda.org & 12,763 \\ \bottomrule
\end{tabular}%
}
\end{table*}

\begin{table*}[]
\centering
\caption{Topics extracted from Stack Overflow questions. The ten topics are categorized into three themes.}
 \label{tab:topics_themes}
 \resizebox{\textwidth}{!}{%
\begin{tabular}{llll}
\toprule
 \textbf{Theme}&\textbf{Topic Id} & \textbf{Topic Name} & \textbf{Sample Keywords}  \\ \midrule
 & 0, 1 & Dependency (2) & dependency, version, specific, release, artifact, resolve, late, update, change, repository   \\  
& 6, 8 & Build (2)  & build, project, create, generate, make, jar, war, compile, failure, resource  \\

Package Management & 3, 11 & Configuration (2)  & package, add, reference, install,  set, environment, variable, import, module, library  \\  
& 7 & Testing & run, test, command, fail, execute, integration, unit, report, clean, surefire   \\ 

& 9, 14 & Error (2) & error, throw, give, exception, load, unable, fix, issue, work, problem   \\  \midrule
 & 10 & Services & application, deploy, web, spring, app, tomcat, deployment, service, engine, boot  \\ 
Input-Output  & 5 & Server-client & server, client, user, request, http, connection, access, response, message, proxy  \\ 
& 2 & Interface & file, time, read, write, image, output, channel, log, process, multiple  \\ \midrule
 \multirow{2}{*}{Package} & 13 & Package function & type, function, struct, string, variable, interface, slice, method, argument, pass  \\  
 & 4, 12 & Package usage (2) & template, react, collection, event, render, helper, database, component, field, document  \\ \bottomrule
\end{tabular}
}
\end{table*}

\begin{figure*}[t]
\centering
     \subfigure[Package ecosystem vs. topic, where users of different package ecosystems ask different questions.]{
       \includegraphics[width=.5\linewidth]{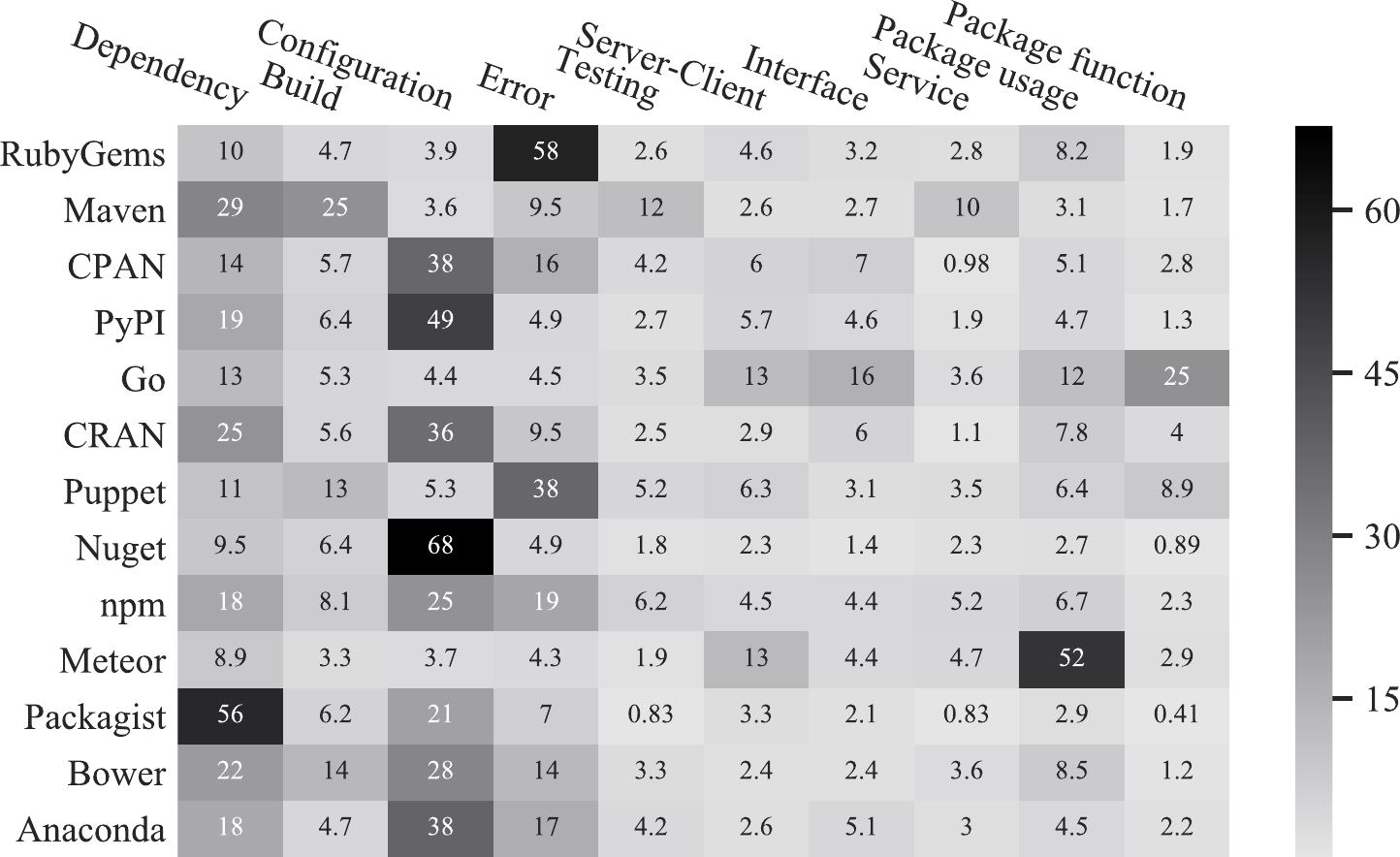}}
        \subfigure[Package ecosystem vs. topic themes, where most users of a package ecosystem ask questions on package management.]{
        \includegraphics[width=.4\linewidth]{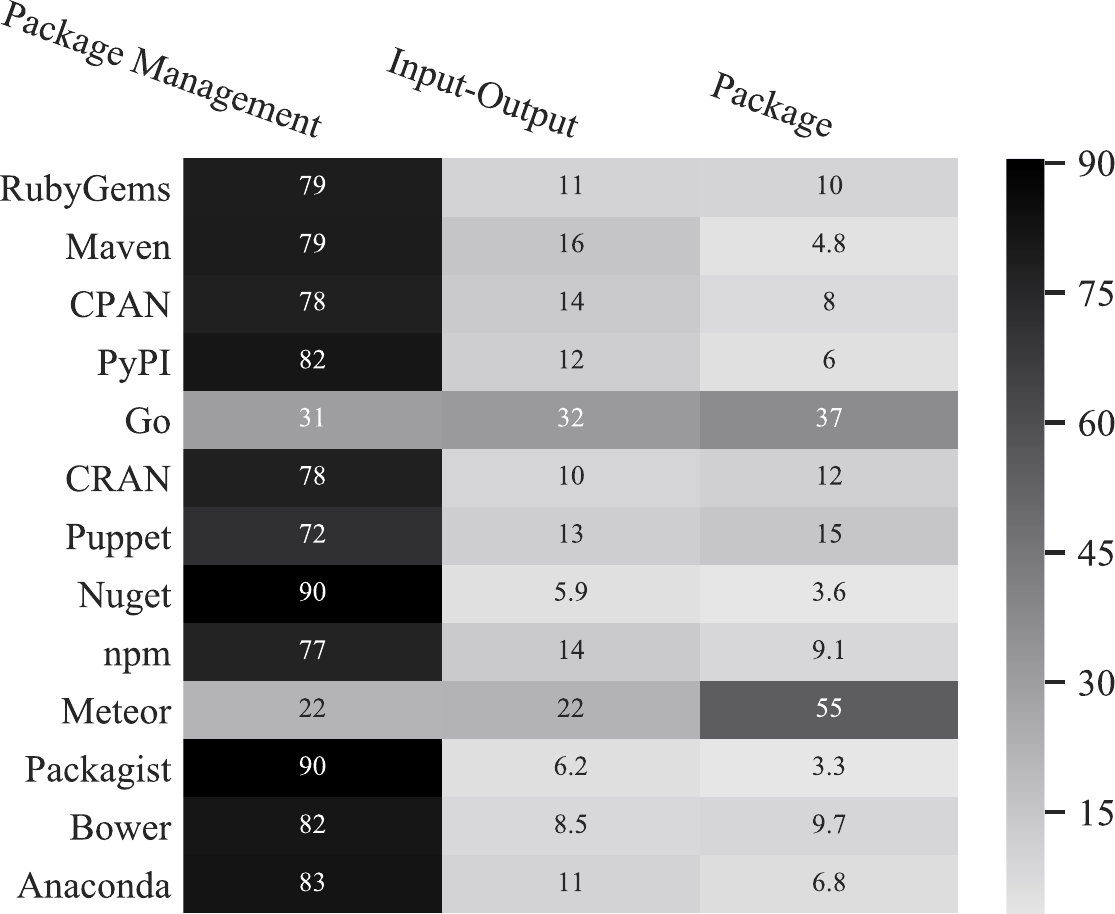}}
       \caption{Heatmaps that show package ecosystem and extracted topics (a) and themes (b). Users of different package ecosystems ask different kinds of questions.}
    \label{fig:topics_themes}
\end{figure*}

The purpose of this paper is to present initial results and a research agenda on how packages are managed within different package ecosystems, and how this impacts the experience of software developers using a package ecosystem. 
We report the results of a preliminary study to analyze developer questions from the commonly used question-and-answer site Stack Overflow.
By mining 497,249 Stack Overflow posts related to 13 different package ecosystems for the 20 top programming languages, we study the topics and difficulties faced by developers.

Our preliminary results show that usage issues can be categorized into ten topics belonging to three themes (i.e., \textit{management}, \textit{input-output}, and \textit{usage}).
Importantly, the results show that the user experience is different, depending on the package ecosystem.
Combining the results, we speculate that users from the Go and Meteor package ecosystems have an easier time finding answers to their questions on Stack Overflow compared to users of other package ecosystems. 

The next logical step in our research agenda is a further exploration of the underlying causes and benefits of using each package ecosystem. 
Our vision is that a series of studies to investigate the benefits and drawbacks from different package ecosystems will help piece together what an ideal package manager and package ecosystem should look like.

\section{Package Ecosystems and their features}
To explore the most popular software ecosystems, we start from the top twenty programming languages from tiobe.com\footnote{Details of the dataset are available at \url{https://www.tiobe.com/tiobe-index/}} as of June,  2021.
To extract the package ecosystems for each language, we use \texttt{libraries.io}\footnote{https://libraries.io/} to gather the package ecosystems that are related to these programming languages.
Note that some of the package ecosystems do not have a dedicated package manager, and that there can be several package ecosystems that are written in one programming language.
Other features include the dependency tree\footnote{An example of differences in dependency trees is described at \url{https://npm.github.io/how-npm-works-docs/npm3/how-npm3-works.html}} and environment.

Table \ref{tab:PM_features} shows a summary of features that are specific to each of our package ecosystems.
In detail, we find that nine of our package ecosystems support a specific programming language (i.e., {\Maven} for Java, {\npm} and {\Bower} for JavaScript, {\Go} for GoLang, {\RubyGems} for Ruby, {\PyPI} for Python, CPAN for perl, {\CRAN} for R, {\Packagist} for PHP).
The other package ecosystems support multiple programming languages.
As shown, a package ecosystem serves from 6,900 to over 1.8 million packages.

\begin{figure}[t]
    \centerline{\includegraphics[width=\linewidth]{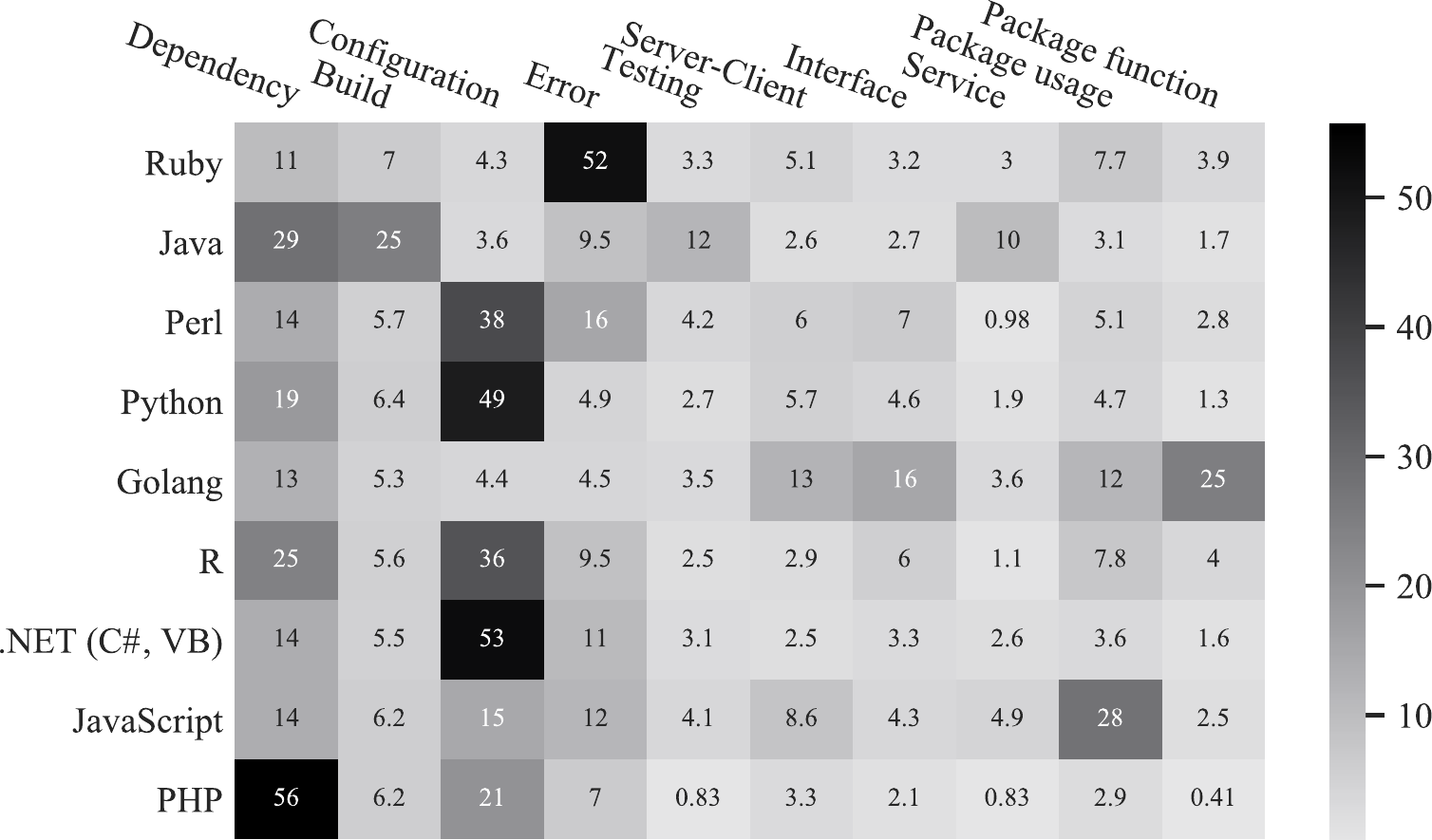}}
    \caption{Language feature heatmap shows that different package ecosystems in different languages provide different user experiences.}
    \label{fig:LangvsTopic}
\end{figure}

\begin{figure}[t]
    \centerline{\includegraphics[width=\linewidth]{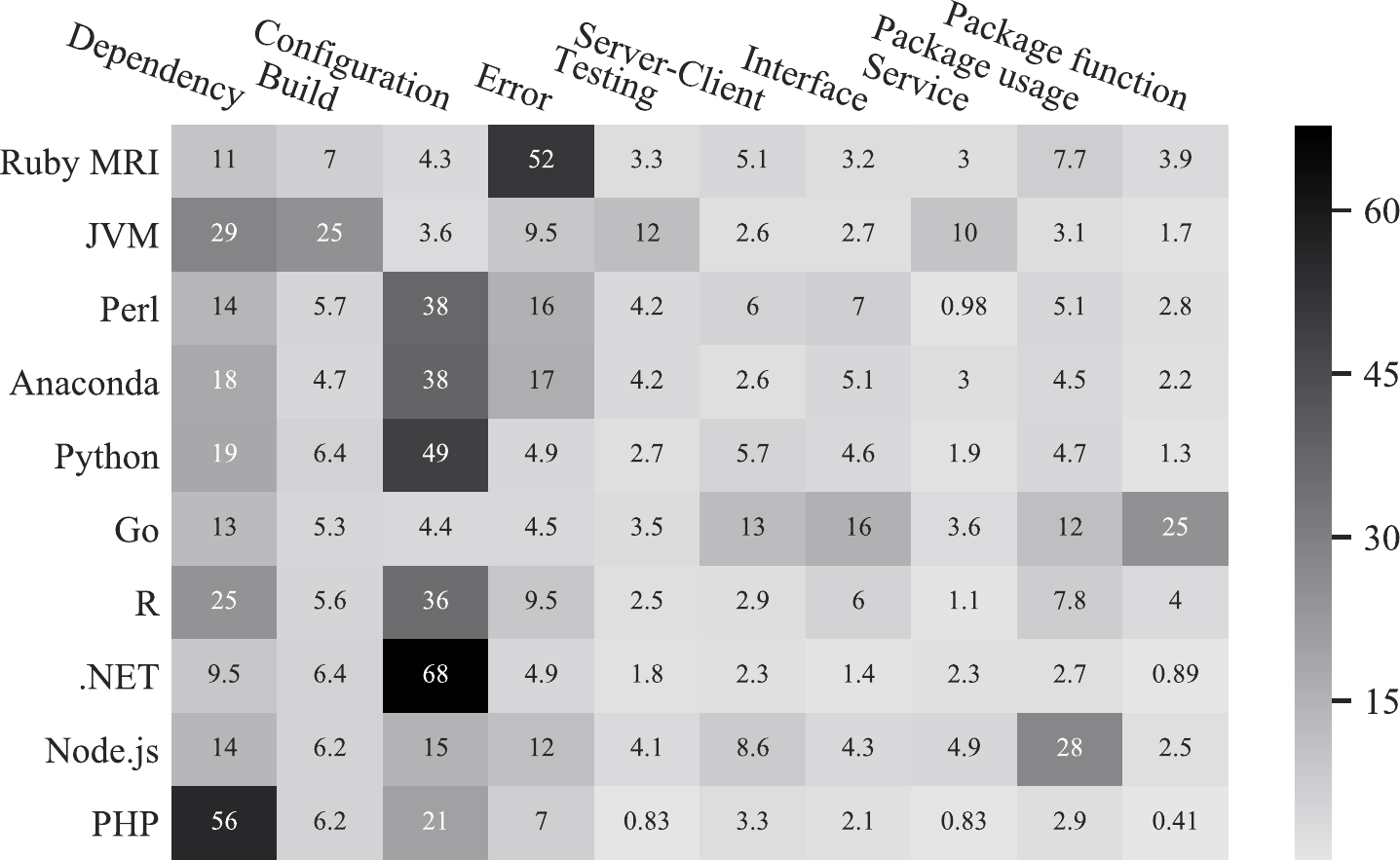}}
    \caption{Environment feature heatmap shows that different package ecosystems in different environments provide different user experiences.}
    \label{fig:EnvvsTopic}
\end{figure}

\begin{figure}[t]
    \centerline{\includegraphics[width=\linewidth]{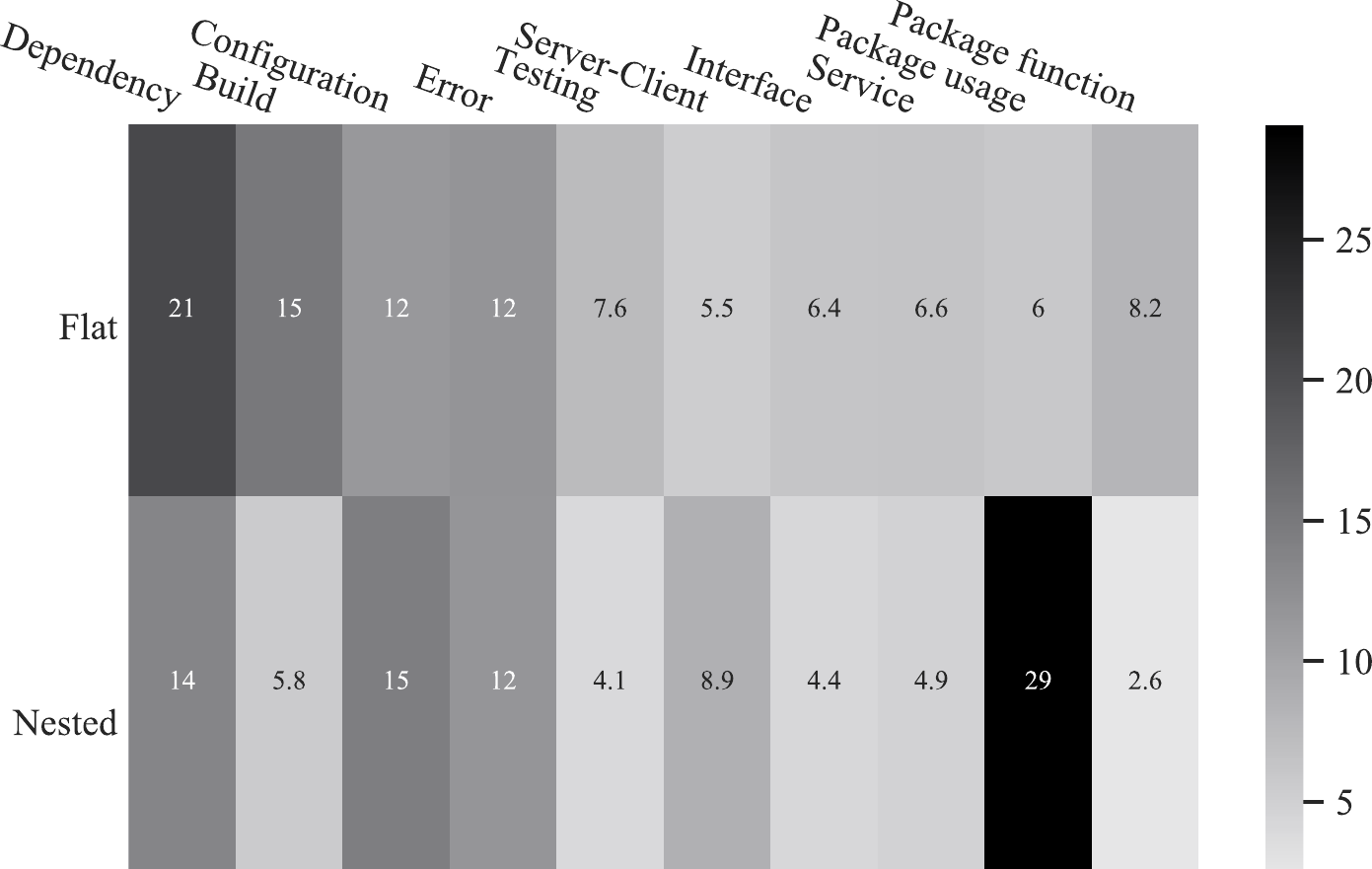}}
    \caption{Dependency tree feature heatmap shows that dependency tree correlates with user experiences (i.e., package ecosystems with a nested dependency face package usage issues, while flat dependency trees face questions related to dependencies)}
    \label{fig:DepTreeTopic}
\end{figure}

\section{Mining User Experience from SO}
To understand the user experience related to different package ecosystems, we mined developer questions about the package ecosystems from Stack Overflow, 

\subsection{Mining Stack Overflow for package manager experience}
We use the SOTorrent dataset~\cite{DBLP:conf/msr/BaltesDT008} as of December 2019. We followed three steps.
In the first step, we used the \texttt{\#package-managers} tag to extract 806 questions.
Second, to discover other relevant tags, we used a semi-automated method to check tags that co-occurred with \texttt{\#package-managers} (i.e., 626 tags).
Three of the authors manually checked each of the 626 tags to determine a list of 28 tags relevant to each package ecosystem.
In the final step, we extracted posts that were tagged with at least one of these 28 relevant tags.
In the end, we collected 497,249 Stack Overflow posts, with 214,609 questions and 282,640 answers.
All data and scripts are made available through our replication package\footnote{Details of the data and scripts: \url{https://doi.org/10.5281/zenodo.5187873}}.

\subsection{Question Topic Modeling}
To explore topics in our dataset, we applied the popular Latent Dirichlet allocation (LDA) topic modeling technique, which is also used in related work (e.g.,~\cite{rosen2016mobile,yang2016security, ahasanuzzaman2018classifying}). 
To prepare our dataset, we first filter out irrelevant information from question titles, namely emails, newline characters, and stop words. We then build a bigram model using Gensim\footnote{Gensim model: \url{https://radimrehurek.com/gensim/}} and lemmatize the words.
Next we use the LDA Mallet (version 2.0.8) model to obtain topics.
We experimented with different topic numbers ($k$) based on coherence score, with $k=15$ giving the best score.

We manually inspected and labelled each topic, resulting in the merge of some of the 15 topics for a total of 10 topics. We also identified three higher-level themes. Table~\ref{tab:topics_themes} shows the three themes and the 10 topics modeled from the user experience.
The first theme is related to package management, which includes keywords such as dependencies, builds, configurations and so on. 
The second theme is related to input and output for packages. This theme includes keywords such as services, server-client and interfaces. 
The final theme is about the low-level topics of package function and package usage.

\section{Preliminary Results}
We contrast developer topic discussions relating to different ecosystems from the two perspectives of (a) package management features, and (b) question popularity and difficulty.

\begin{table*}[t]
\centering
\caption{Summary of popularity and difficulty of topics. }
 \label{tab:topicsPD}
\resizebox{0.7\textwidth}{!}{%
\begin{tabular}{@{}llrcrcc@{}}
\toprule
 &  & \multicolumn{3}{c}{Popularity} & \multicolumn{2}{c}{Difficulty} \\
Theme & Topic & \# Questions & \begin{tabular}[c]{@{}c@{}}\#Score \\ (median)\end{tabular} & \begin{tabular}[c]{@{}c@{}}\#Views \\ (median)\end{tabular} & \begin{tabular}[c]{@{}c@{}}Questions with accepted \\ answer (\%)\end{tabular} & \begin{tabular}[c]{@{}c@{}}PD Score\\ (\%)\end{tabular} \\ \midrule
 & Dependency & 40,519 & 1 & 420 & 48 & 0.24 \\
 & Build & 27,191 & 1 & 409 & 46 & 0.24 \\
Package Management & Configuration & 26,748 & 1 & 348 & 45 & 0.29 \\
 & Error & 25,731 & 1 & 447 & 44 & 0.22 \\
 & Testing & 14,194 & 1 & 440.5 & 45 & 0.23 \\ \midrule
 & Server-Client & 13,770 & 1 & 319 & 48 & 0.31 \\
Input-Output & Interface & 12,582 & 1 & 277 & 53 & 0.36 \\
 & Services & 13,256 & 0 & 405 & 43 & 0.25 \\ \midrule
\multirow{2}{*}{Package} & Package usage & 26,226 & 0 & 216 & 53 & 0.46 \\
 & Package function & 14,392 & 1 & 284 & 64 & 0.35 \\ \midrule
Total &  & 214,609 &  &  &  &  \\ \bottomrule
\end{tabular}%
}
\end{table*}

\subsection{Contrasts in Features: Topics and Features}
We use heatmap visualizations to contrast differences between the features of a package ecosystem and the kinds of topics that users ask about on Stack Overflow.
The heatmap uses greyscale colored cells to show a two-dimensional matrix between the topics and the features of the package ecosystem.
We report the frequency counts of each dimension that is reflected in the cells.

Figure~\ref{fig:topics_themes} shows that the topics related to each package ecosystem differ.
For example, package ecosystems like CRAN, CPAN, and Conda tend to attract configuration related questions, while Go and Meteor have their users ask questions related to the package function and usage. 
Under the broader themes, the results are consistent with the results of Table \ref{tab:topics_themes}, as most topics are related to package management.
This evidence suggests that users of Go and Meteor package ecosystems may face different types of issues when compared to users of other package ecosystems.

\begin{tcolorbox}
\textbf{Takeaway 1:}
Users from different package ecosystems report different issues.
Findings indicate that RubyGem users report errors, while NuGet users report configuration issues. 
\end{tcolorbox}

Taking a deeper look at the features, we can see that users of package ecosystems built for JavaScript technologies and environments tend to encounter different types of issues, whereas users of package ecosystems built for the Python, R and Perl languages and related environments tend to focus on configuration issues (cf. Figure~\ref{fig:LangvsTopic} and Figure~\ref{fig:EnvvsTopic}).
In terms of the dependency tree, Figure~\ref{fig:DepTreeTopic} shows that flat dependencies tend to attract dependency related issues, while users of package ecosystems with nested dependency trees encounter package usage related issues.
One possible explanation is that nested dependencies are a solution to `dependency hell', thus removing dependency related issues for package ecosystems with nested dependency trees\footnote{This issue is discussed by this npm blog post at \url{https://npm.github.io/how-npm-works-docs/theory-and-design/dependency-hell.html}}.

\begin{tcolorbox}
\textbf{Takeaway 2:}
Users from different programming languages report different kinds of issues. 
We find that users of applications that are developed using the Python language reported more configuration issues, when compared to Ruby users who report errors.
\end{tcolorbox}

\subsection{Contrasts in Responses: Popularity and Difficulty}
To characterize package ecosystem topics in terms of their popularity and difficulty, we measure both metrics as defined by Yang et al. \cite{yang2016security}.
We calculate the PD (post difficulty) score as $\frac{AnswerCount\ (median)}{ViewCount\ (median)}$ for each post, i.e., the smaller the PD score, the more difficult it is to answer a question. We then summarize the median PD (i.e., $PD(median)$) for each topic and the median value for the popularity (i.e., $Popularity(median)$).

Table~\ref{tab:topicsPD} shows that errors and testing are the most difficult topics to provide answers for.
In contrast, questions that relate to package usage and package function are the easiest for users to get an answer. 
Combining these results with the relationships between themes and package ecosystems (cf.~Figure \ref{fig:topics_themes}), we speculate that users of the Go and Meteor package ecosystems may have a relatively easy experience in finding answers to their questions on Stack Overflow. 

\begin{tcolorbox}
\textbf{Takeaway 3:}
Different topics imply different degrees of difficulty.
Errors and testing are the most difficult, while package usage is the easiest to answer.
Combining with Takeaway 1, users of the Go and Meteor package ecosystems face a relatively easy experience.
\end{tcolorbox}

 \section{Limitations}
The first threat to internal validity is related to our collected data.
We acknowledge that some posts may be mislabelled (i.e., missing tags or incorrect tags) on Stack Overflow.
The second threat is related to the correctness of techniques used in this study, such as choosing the appropriate number of topics (k=15) for the LDA model. A different number might have led to different results. Manually labeling the topics based on keywords introduced another threat to the validity. 
We mitigated this threat by involving multiple authors in the labelling process.

\section{Implications}
We now discuss implications of our work applicable to developers and package manager designers:

\paragraph{Developers}
Developers should be conscious that their choice of a package ecosystem will impact their user experience -- not all package ecosystems are the same as design choices such as hierarchy structure and language support are correlated with what questions developers are likely to encounter. When starting a new project, developers might be able to choose an ecosystem based on our insights. In other cases it might be too late to switch ecosystems, but users might still be able to consider the benefits of (somewhat) compatible package ecosystems, such as npm and Meteor.
Furthermore, as applications such as mobile apps require support for multiple technology stacks, developers should be aware of the trade-offs when switching technologies, i.e., when porting a Java application (\Maven) to the web as a JavaScript application (\npm).
Developers can use our findings to better understand what technical background knowledge they should have regarding package ecosystems.

\paragraph{Package manager designers}
Designers need to be aware of the impact of package ecosystem design features on the problems that developers may encounter as some design choices are correlated with more questions on certain topics raised on Stack Overflow.
Designers should be proactive about issues frequently encountered by package ecosystem users, by providing thorough documentation and/or improving package ecosystems where possible.
Designers should also make it easy for developers to find the information they need to resolve issues, e.g., by providing good error messages, while other package ecosystems may require better documentation for package manager configuration.

\section{Roadmap}
In this preliminary study, we explore thirteen package ecosystems to understand whether their differences in terms of features correlate with user experience.
We find that developers ask package ecosystem related questions that can be clustered into ten different topics, which can be grouped into three themes, and that different topics are prevalent for different ecosystems.

The next logical step is further exploration into the underlying causes and benefits when using each package ecosystem. 
A thorough study to investigate the trade-offs between benefits and drawbacks of each package ecosystem will help piece together what the ideal package ecosystem should look like. Researchers could use our findings to prioritise research efforts, as our work is the first to acknowledge that developers encounter issues when using package ecosystems.

Future work is needed in particular in teasing apart the effects of different features of package ecosystems and their interplay. While some features of a package ecosystem are a given (e.g., the programming language), others provide more freedom to designers (e.g., dependency trees). In addition, interviewing, observing, and/or surveying software developers would further allow for triangulation of our preliminary findings to understand the impact of ecosystem features on developers' decision processes.

\textbf{Acknowledgments.} 
This work is supported by Japanese Society for the Promotion of Science (JSPS) KAKENHI Grant Numbers JP18H04094, JP20K19774, and JP20H05706.


\end{sloppy}
\end{document}